\documentclass[twocolumn,prd,nofootinbib,superscriptaddress,eqsecnum,tightenlines,8pt]{revtex4}

\usepackage{fancyhdr}
\usepackage{amsmath,amsfonts,amssymb}
\usepackage[colorlinks,citecolor=blue,linkcolor=blue,urlcolor=blue]{hyperref}
\usepackage[english]{babel}

\usepackage{color}
\definecolor{red}{rgb}{1,0,0}
\definecolor{gre}{rgb}{0,0.6,0}
\definecolor{blu}{rgb}{0,0,1}

\usepackage{mathenv}
\usepackage{ulem}

\usepackage[pdftex]{graphicx}
\usepackage{mathrsfs} 
\def\be{\begin{equation}}
\def\ee{\end{equation}}

\newcommand{\nn}{\nonumber}
\renewcommand{\d}{\textrm{d}}

\newcommand{\su}{\mathfrak{su}(2)}

\newcommand{\p}{\partial}

\newcommand{\gb}{\bar{g}}

\newcommand{\cR}{\mathcal{R}}

\newcommand{\Ls}{M}

\usepackage{color}

\def\del {\partial}

\def\Z{{\mathbb Z}}

\def\R{{\mathbb R}}

\def\mQ{\mathcal{Q}}

\def\f{\frac}

\def\d{\dot}

\def\bra{\langle}
\def\ket{\rangle}
\def\dd{{\rm d}}

\def\del{\partial}

\def\vp{\varphi}

\def\mR{\mathcal{R}}

\def\cA{{\mathcal A}}

\def\cC{{\mathcal C}}

\def\cH{{\mathcal H}}

\def\cR{{\mathcal R}}
\def\cS{{\mathcal S}}

\def\Tr{{\rm Tr}}

\def\cancel#1#2{\ooalign{$\hfil#1\mkern1mu/\hfil$\crcr$#1#2$}}
\def\Dirac{\mathpalette\cancel D}

\begin{document}

\title{Testing different approaches to quantum gravity with cosmology: An overview}

\date{\today}

\author{Aur\'elien Barrau}
\email{barrau@in2p3.fr}
\affiliation{
Laboratoire de Physique Subatomique et de Cosmologie, Universit\'e Grenoble-Alpes, CNRS-IN2P3\\
53, Avenue des Martyrs, 38026 Grenoble cedex, France\\
}%

\begin{abstract}
Among the available quantum gravity proposals, string theory, loop quantum gravity, non-commutative geometry, group field theory, causal sets, asymptotic safety, causal dynamical triangulation, emergent gravity are among the best motivated models. As an introductory summary to this special issue of {\it Comptes Rendus Physique}, I explain how those different theories can be tested or constrained by cosmological observations. 
\end{abstract}

\maketitle

\section{Introduction}

A century after the first claim, by Einstein, that general relativity needs to be quantized, a consensual solution is still missing. There are however several convincing  approaches available. Obviously, small black holes and the early Universe are the best ``places" to confront those models with observations. As light black holes have never been observed, primordial cosmology is probably the most promising situation to consider. I try to review here the merits and weaknesses of each approach when compared with cosmological observations. This article does not pretend to be exhaustive or unbiased and is highly based on the invited articles written for the special {\it Comptes Rendus Physique} issue ``testing quantum gravity with cosmology", that I coordinate. \\

For some quite well known approaches, I focus only on the cosmological consequences. For other, less konwn ones, I also recall the basics of the construction.\\

Relevant references for most of the models described in the following are given in the associated articles published in this special volume. The reference list given at the end of this article is therefore in no way sufficient.

\section{String theory}

String theory is a fascinating proposal in which quantum strings (instead of usual point particles) propagate in space and interact with each other. This approach might unify all particles and interactions and solve the quantum gravity problem. It is notably hard to test and one might argue that it is a framework rather than a clearly defined theory. It is however fair to mention that predictions can be made and confronted with observations. As an example, a detection of primordial gravitational waves in the near futur might put string theory under serious pressure \cite{Parameswaran:2016fqr}. 

An even simpler path (at the conceptual level) was followed in \cite{Andriot:2016xvq} focusing on the possibility to find a ``well understood" (that is with a correct 10-dimensional description) metastable de Sitter vacuum in string theory. This is arguably something to expect from the theory, so as to be able to describe both the inflationary period and the contemporary acceleration. The main idea to construct a de Sitter vacuum is to use anti-branes to uplift the value of the cosmological constant, but the full D=10 solutions are not well understood. Many no-go theorems on classical dS solutions were derived. Let me take a specific example. Without using non-geometric fluxes, the study performed in \cite{Andriot:2016xvq} relies on D=10 type II supergravities -- describing space-time as the warped product of a 4-dimensional dS one with a 6-dimensional compact internal manifold -- ignoring $\alpha'$ corrections, supplemented by the Ramond-Ramond sources $D_p$-branes and orientifold $O_p$-planes. This is especially relevant considering recent constrains on dS vacua in supersymmetric heterotic string theory. As a result, is was shown that the cases $p=3$, $7$ or $8$ are strictly excluded whereas severe constraints can be derived for $p=4$, $5$, $6$. This is basically achieved thanks to a derivation of the 4-dimensional Ricci scalar. Typically, using the Bianchi identities projected in the transverse direction, one obtains

\begin{eqnarray}
{\cal R}_4 + 2 (\nabla\del \phi)_4  &=& -\frac{2}{p+1} \bigg( -4 |\del \phi|^2 + 2 \Delta \phi -  2 \varepsilon_p e^{\phi} (\d F_k)_{\bot}  \nn\\
&+& \left|*_{\bot}H|_{\bot} + \varepsilon_p e^{\phi} F_{k-2}|_{\bot} \right|^2 + |H|^2 - |H|_{\bot}|^2  \nn\\
&+&e^{2\phi} (  |F_{k-2}|^2 - |F_{k-2}|_{\bot}|^2 + 2 |F_{k}|^2) \nn\\
&+& 3 |F_{k+2}|^2 + e^{2\phi} ( 4 |F_{k+4}|^2 + 5 |F_{k+6}|^2 )  \bigg)  \ . \nn
\end{eqnarray}

The only term in the right-hand side with indefinite sign being, in the smeared limit, $(\d F_k)_{\bot}$, this establishes that there is no dS vacuum in the limit $(\d F_k)_{\bot}\rightarrow 0$. This line of research is very promising.\\

Let me now go back to the basics to introduce the approach presented in this special volume. String theory compactification leads to massless scalars (moduli) that are not observed. Such moduli are of two kinds: complex structure moduli and K\"ahler moduli. The mechanism to fix the moduli and solve this problem consists in adding background p-form fluxes (10-dimensional generalizations of the electromagnetic flux) which wrap p-cycles of the compactification manifold. The first kind of moduli can be fixed by a specific combination of 3-form fluxes in type IIB string theory so that the back-reaction on the geometry still allows for a Calabi-Yau manifold \cite{Grana:2000jj}.  The second kind was fixed by using  non-perturbative quantum corrections to the relevant action \cite{Kachru:2003aw}. At this stage, all solutions are still AdS and the last part of the game consists in uplifting the value of the cosmological constant. Since there exist large number of Calabi-Yau manifolds, and a very large number of possibilities of putting flux on them, it can be shown that there might exist of order $10^{500}$ flux compactifications and as many (effective) physical laws and (nearly) fundamental constants. As explained by Bena and Gra\~na in this volume \cite{bena} this could lead to a drastically new view of physics in which one considers the constants we measure in our Universe as environmental (anthropic) variables in a larger Multiverse (see, {\it e.g.} \cite{Barrau:2007ce}). They argue that two paradigms are now competing: the {\it Theory of Everything}  and the {\it Anthropic/Multiverse}. Those ideas are not intrinsically related with string theory but string theory might allow to address the discrepancy between these two approaches by using controlled calculations.\\

The {\it Anthropic/Multiverse paradigm} offers an easy (which does not mean correct) way to account for three recent  results:
\begin{itemize}
\item the cosmological constant, which is 120 orders of magnitude smaller than the value predicted by quantum field theory (it should be noticed that one could also argue that a ``real cosmological constant" has nothing to do with particle physics and that the question to understand why quantum fluctuations do not gravitate is a different one). 
\item the hierarchy problem: there are 24 orders of magnitude between the electroweak scale and the Planck scale. Why is the Higgs mass so light whereas quantum correction should increase the mass close to the Planck scale.
\item the measured flatness of the inflationary potential might also appear as an issue (although a less severe one).
\end{itemize}

The interesting novelty is that, as underlined by the example I gave first, many no-go theorems have been derived for constructing dS vacua. There are two main approaches to derive viable dS solutions: either by directly constructing classical non-supersymmetric dS spaces, or by building supersymmetric flux compactifications with all the moduli stabilized, and then uplifting the negative cosmological constant to a positive value. For many reasons discussed in details in \cite{bena} obtaining dS vacua in String Theory is difficult and, beyond the construction itself, there is still the severe issue of  stability. Some instabilities are obvious ({\it e.g.} those arising in universal sectors, namely because of scalars appearing in gauged supergravity coming from a string compactification: the dilaton, the overall volume, or the Goldstino direction in solutions where SUSY is spontaneously broken) but others are hidden (due to fields outside of the truncation performed to obtain the effective action).

Stability issues are serious ones. Even in the enormous landscape of AdS vacua, looking at the truncation shows (using the weak gravity conjecture), that all non-supersymmetric AdS solutions could be in the space of consistent-looking semiclassical effective theories which are actually inconsistent, with instabilities coming from an interplay between closed an open string sectors. It is therefore worth stressing that tremendous progresses are being made in trying to understand wether the actual cosmological background space can be obtained (or not) in string theory. This leads to important constraints on the theory and could even put it under pressure. It is both fascinating and intriguing that the apparently easy task to accommodate for a positive cosmological constant is so hard in string theory.

Progresses are slower in making clear and universal predictions for the early universe, in particular for the observed spectra.

\section{Loop quantum gravity}

Loop quantum gravity (LQG) is a tentative non-perturbative and background independent quantization of general relativity. It can be expressed both in a canonical way and in a covariant formalism (spinfoams). 
In loop quantum cosmology (LQC), symmetry-reduced geometries are quantized using the same procedures as in LQG. The main achievement of LQC is that the big-bang and big-crunch singularities are solved by quantum gravity effects. Instead, a non-singular bounce takes place. 

The geometrical sector of the phase space can be described by the $\su$-valued Ashtekar connection $A_a = A_a^i \tau_i$ and the associated conjugate momentum, the densitized triad $E^a_i$. One then defines the holonomy of the  Ashtekar connection and the flux of the densitized triad and the full kinematical Hilbert space can be constructed in a rigorous and well defined way. \\

The LQC quantum dynamics has been studied choosing an initial state $\Psi(V, \phi_o)$ at some $V = a^3$ and some moment $\phi_o$, and evolving it using the Hamiltonian constraint operator.  This was first performed for sharply peaked initial states, numerically solving the dynamical equations. This shows that the wave function remains sharply peaked throughout the evolution, that the wave packet follows the classical trajectory very closely when the energy density is small, and that, when the energy density approaches the Planck scale, it departs from the classical theory and bounces at $\rho_c \sim \rho_{\rm Pl}$.  Many recent studies have shown that generic widely spread states (with no semi-classical limit) also bounce.  In addition, the quantization ambiguities in the Hamiltonian constraint operator have also been investigated, showing that the bounce occurs anyway.\\

For sharply peaked states the LQC effective Friedmann equations reads:
\be \label{fr1}
H^2 = \f{8 \pi G}{3} \rho \left(1 - \f{\rho}{\rho_c}\right),
\ee
\be \label{fr2}
\dot{H} = -4 \pi G (\rho + P) \left(1 - \f{2 \rho}{\rho_c}\right).
\ee

The bounce clearly takes place when $\rho = \rho_c$.  It should also be noticed that while the equations of motion for gravity are modified by LQG corrections, the continuity equation $\dot\rho + 3 H (\rho + P) = 0$ for the matter sector remains unchanged.\\

There are three leading approaches to cosmological perturbation in LQC. The first one is called ``effective constraints" or ``deformed algebra". The idea is to try to find the correct effective equations without having the full underlying quantum theory.  One takes the classical scalar and diffeomorphism constraints for the FLRW space-time with linear perturbations, and then allows for possible modifications motivated by LQC (holonomies or inverse triads). Some freedom appearing in the choices of the `correction' functions is removed by the condition -- necessary to obtain a consistent theory -- that the constraints exhibit an anomaly-free Poisson algebra. For scalar perturbations, the holonomy-corrected Mukhanov-Sasaki equation  is 
\be \label{ef-sc}
v_k'' + \left(1 - \f{2 \rho}{\rho_c}\right) k^2 v_k - \f{z_s''}{z_s} v_k = 0,
\ee
and for tensor perturbations, is becomes
\be \label{ef-te}
\tilde \mu_k'' + \left(1 - \f{2 \rho}{\rho_c}\right) k^2 \tilde \mu_k - \f{z_t''}{z_t} \tilde \mu_k = 0,
\ee
where $\tilde \mu_k$ is related to $h_k$ by $\tilde \mu_k = z_t h_k$ and $z_t = a / \sqrt{1 - 2 \rho / \rho_c}$.
In addition, the constraint algebra leads to some interesting speculations concerning a possible change of signature around the bounce.
The main drawback of this approach in that it ignores quantum fluctuations and cannot be trusted to evolve trans-Planckian modes through the bounce.\\

The second framework developed for cosmological perturbation theory in LQC is called ``hybrid quantization" or ``the dressed metric" approach. The idea of the hybrid quantization is to treat the background and perturbative degrees of freedom differently, by performing an LQG-like quantization of the background and a Fock-like quantization of the perturbations. Importantly, quantum fluctuations, that play an important role for trans-Planckian modes, are taken into account. 
This gives a fully quantum treatment of  perturbations on a quantum background.
The equations of motion (scalar perturbations) in the dressed metric approach are given by
\be \label{sc-hybrid}
\hat\mQ_k'' + 2 \f{\tilde a'}{\tilde a} \hat\mQ_k' + \left( k^2 + \f{\tilde a''}{\tilde a} - \f{\tilde u''}{\tilde u} \right) \hat \mQ_k = 0,
\ee
with $\mQ = v / a = z_s \mR / a$, $u = a \sqrt{3 (1 + w_{eff}) / 8 \pi G}$,  $w_{eff} = [\dot\phi^2 / 2 - V(\phi)] / [\dot\phi^2 / 2 + V(\phi)]$, and
\be
\tilde a^4 = \f{\bra \hat H_o^{-1/2} \, \hat a^4 \, \hat H_o^{-1/2} \ket}{\bra H_o^{-1} \ket},
\ee
\be
\tilde u = \f{\bra \hat H_o^{-1/2} \, \hat a^2 \, \hat u \, \hat a^2 \, \hat H_o^{-1/2} \ket}
{\bra \hat H_o^{-1/2} \, \hat a^4 \, \hat H_o^{-1/2} \ket},
\ee
giving the required expectation values. $\hat H_o$ is the  LQC Hamiltonian for the background with respect to the time variable $\phi$. In this approach, initial conditions are usually imposed at the bounce point for heuristic reasons.\\

The third approach is called the ``separate universe approximation". This framework was adapted to LQC to provide a consistent loop quantization for both the background and long-wavelength perturbations. The main idea is to discretize a space-time with small perturbations into a lattice. Each cell in this lattice is approximately homogeneous.  Only long-wavelength modes are included, cells are therefore uninteracting with each other and a loop quantization is possible in each cell. This quantization is simple for scalar perturbations in the so-called longitudinal gauge. Effective equations can be used for each cell to capture the dynamics.  The equations of motion for perturbations are given by (for sharply peaked states):
\be \label{sep-v}
v_k'' - \f{z_s''}{z_s} \, v_k = 0.
\ee
The separate universe approach has the advantage that it is the only one that allows for a LQG quantization of both the background and perturbations.  However it is only applicable to long-wavelength scalar perturbations and requires a gauge-fixing before quantization.\\

The predictions of LQC, just like in usual general relativity, do depend on the matter content, and therefore LQC effects will change depending on the details of the considered scenario.  In addition, since the three approaches to cosmological perturbations in LQC outlined above have some differences, the predictions may also depend on the specific chosen approach. At this stage it is therefore difficult to draw fully general conclusions. It seems however that a quite generic prediction is a power spectrum of curvature perturbations with less power at large scales if the number of inflationnary e-folds is close to its minimum allowed value. The same power suppression effect is also predicted in the T-E and E-E correlation functions as well as in the B-mode power spectrum. Some ``universal" LQC predictions are slowly being understood \cite{Bolliet:2015bka}.

\section{Non-commutative geometry}

The spectral action functional  \cite{Chamseddine:1996zu}  is a proposal for gravity, coupled to matter, on noncommutative spaces. Physics on a compact Riemannian smooth manifold is generalized to the noncommutative world by the so-called spectral triples $(\cA,\cH,D)$, which contains the relations between the algebra of functions
$\cA=\cC^\infty(X)$ and the metric (encoded in the Dirac operator $\Dirac$ acting on the Hilbert space $\cH=L^2(X,S)$) on the manifold $X$. The commutators $[D,a]$
are bounded and are acting on $\cH$ while $D$ is required to be self-adjoint. The spectral action is a kind of regularized trace of the Dirac operator. Assuming that the
operator $D$ is such that $\Tr(|D|^{-s})<\infty$ for large enough $Re(s)$, the spectral action functional reads as 
\begin{equation}
\cS_{\Lambda,f}(D) = \Tr ( f(D/\Lambda) = 
\sum_{\lambda \in \text{Spec}(D)}{\rm Mult}(\lambda)f(\lambda/\Lambda) , 
\end{equation}
where $f\in \cS(\R)$ is a kind of smooth cutoff function and $\Lambda\in \R^+$ is the energy scale that makes $D/\Lambda$ dimensionless.

This allows the construction of particle physics models, where the asymptotic expansion coincides with the Lagrangian of the Standard Model. In the gravity sector,
the asymptotics of the spectral action leads to modified gravity where, in addition to the Einstein--Hilbert action and the cosmological
constant, the Lagrangian also contents Weyl Curvature terms (conformal gravity) and the Gauss--Bonnet gravity term (which is a topological in D=4).\\

As explained by Marcolli in this volume \cite{marcolli}, this model has interesting cosmological consequences. The asymptotic expansion of the spectral action can be written as

\begin{widetext}
\begin{eqnarray}{rl}
 \cS_{\Lambda,f}(D) \sim & \displaystyle{\frac{1}{2\kappa_0^2(\Lambda)}  \int\,R
 \, \sqrt g \,d^4 x + \gamma_0(\Lambda) \,\int \,\sqrt g\,d^4 x + \displaystyle{ \alpha_0(\Lambda)\, \int C_{\mu
\nu \rho \sigma} \, C^{\mu \nu \rho \sigma} \sqrt g \,d^4 x + \tau_0(\Lambda)\, \int R^* R^* \sqrt g \,d^4 x}} \\[3mm]
     +& \displaystyle{ \frac{1}{2} \int\,  |D H|^2\, 
\sqrt g \,d^4 x -  \mu_0^2(\Lambda)\, \int\,  |H|^2\, \sqrt g \,d^4 x } - \displaystyle{ \xi_0(\Lambda)\,  \int\, R \, |H|^2 \, \sqrt g \,d^4 x
+ \lambda_0(\Lambda)\,  \int |H|^4 \, \sqrt g \,d^4 x }
\\[3mm]
 +& \displaystyle{ \frac{1}{4} \int\,(G_{\mu \nu}^i \, 
G^{\mu \nu i} +  F_{\mu
\nu}^{ \alpha} \, F^{\mu \nu  \alpha}+\, B_{\mu \nu} \, B^{\mu \nu})\, \sqrt g \,d^4 x },
\end{eqnarray} 
\end{widetext}

where $H$ is the Higgs, $G,F,B$ are gauge bosons, $C_{\mu\nu\rho\sigma}$
is the Weyl curvature, and $R^*R*$ is the Gauss--Bonnet term. The coefficients correspond to  an ``effective cosmological
constant" and to an ``effective gravitational constant" (functions of the Yukawa parameters and of the Majorana mass
matrix, which in turn run with the energy scale $\Lambda$).\\

Unlike the Einstein--Hilbert action, which is sensitive only to the local curvature but not to the topology, the spectral action depends (through the spectrum of the Dirac operator) on global properties. Interestingly, it was shown that the  cosmic topology can leave a signature on the shape of the inflaton potential in spectral action cosmology. This will impact CMB scalar and tensor power spectra. It was even suggested that, if the spectral action is calculated on an almost commutative geometry, the potential
acquires a multiplicative factor that depends on the number of fermionic  particles in the matter sector of the model.  

It is also particularly interesting to investigate spectral action consequences for non-homogeneous
spacetimes like the Packed Swiss Cheese Cosmologies derived from an Apollonian packing of
$3$-spheres in a D=4 spacetime. This leads to multifractal cosmologies with possible explanations for the large scale distribution of galaxies.
Under some reasonable assumptions, the leading terms in the spectral action expansion takes the form
\begin{equation} 
\cS_{\Lambda,f}(D) \sim \Lambda^\sigma \sum_{m\in \Z} \Lambda^{\frac{2\pi i}{\log b}} f_{s_m},
\end{equation} 
where $s_m =\sigma+ \frac{2\pi i m}{\log b}$
are the poles of the zeta function and $\sigma=\frac{\log b}{\log a}$ is the Hausdorff dimension.
An inflaton potential $V(\phi)$ can then be obtained and an additional term appears, which changes the slow-roll parameters and the indices of the scalar and
tensor fluctuation. In principle, signatures of the multifractality of the
spacetime structure could be measured but no smoking gun observable has yet been identified.

\section{group field theory}

Group field theories (GFTs) are quantum field theories on a group manifold, characterized by specific non-local interactions. 

As explained by Oriti \cite{Oriti:2016acw} in this volume, the variable of GFT is a complex field $\varphi \,:\, G^{\times d} \rightarrow \mathbb{C}$, where $G$ is a Lie group. In GFT, the fields should not be understood as having values in a spacetime manifold. The phase space for a ``quantum" of the GFT field (a kind of atom of space) is $\left( \mathcal{T}^*G\right)^d$, and the associated Hilbert space is $\mathcal{H} \,=\,L^2\left( G^d\right)$. The fields $\varphi(g_i)=\varphi(g_1,...,g_d)$ can be expressed in terms of dual Lie algebra variables through a non-commutative Fourier transform or, alternatively, as irreducible representations of $G$. For D=4 (Lorentzian) quantum gravity, the correct group is the Lorentz group $SO(3,1)$ or its rotation subgroup $SU(2)$.  A quantum can be seen as a topological polyhedron with faces labelled by the arguments of the field. The usual choice are 3-simplices (that is tetrahedras). The group, Lie algebra or variables labelling GFT states can be interpreted as discrete connection or metric variables, representing the geometry of the  polyhedral structures. 
The group field theory kinematical Hilbert space is such that 
\begin{equation} 
\mathcal{F}\left( \mathcal{H}\right) \, =\, \bigoplus_{N=0}^{\infty} sym \left\{ \mathcal{H}^{(1)} \otimes \cdots  \mathcal{H}^{(N)}\right\}\qquad \mathcal{H} \,=\,L^2\left( G^{\times d}\right) \qquad ,
\end{equation} 
where a bosonic statistics is assumed and ladder field operators are used
$\hat{\varphi}(g_{1},g_{2},g_{3},g_{4}), \hat{\varphi}^\dagger(g_{1},g_{2},g_{3},g_{4})$.

The dynamics of  GFT can be derived from an action which includes quadratic terms and higher order interaction ones. Group field theories are special because the pairing of field arguments in the interaction part is non-local in the sense that they are  matched to one another only according to specific patterns whose combinatorics is included of the definition of the theory. Usually the only interaction terms are five-valent: 
\begin{eqnarray}
S_\lambda &=& \int  \dd g_{v_1} \dd g_{v_2}  \bar\vp(g_{v_1}) \vp(g_{v_2}) \, K_2\\
&+& \f{\lambda}{5} \int \left(\prod_{a=1}^5 \dd g_{v_a} \bar\vp(g_{v_a}) \right) \bar{\mathcal{V}}_5\\
&+& \f{\lambda}{5} \int \left(\prod_{a=1}^5 \dd g_{v_a} \vp(g_{v_a}) \right) \mathcal{V}_5.
\end{eqnarray}

Recovering the usual space and time from the discrete and algebraic structure of GFT is challenging.  In a way  the real degrees of freedom of the theory are non-spatiotemporal, and geometry is {\it emergent}. Recently, a new direction was explored, taking into account more and more degrees of freedom and moving from the regime in which only simple spin network graphs were used to an approach using the renormalization group. One of the main goals is to identify continuum phases. This has led to the idea of GFT condensate cosmology. It consists in looking for the general relativistic dynamics as the hydrodynamics approximation of the fundamental theory, and for cosmology in the most coarse grained sector of the same effective theory. \\

One can expect to have a description of the macroscopic universe as a fluid, whose atoms  are the GFT fundamental quanta, and whose main collective variable is a density function, plus a velocity function \cite{Gielen:2013kla}. As the general problem is outstandingly complicated, one focuses on the case of quantum condensates or superfluids. The quantum states are formed by an infinite number of GFT quanta, and they encode the information corresponding to the phase space of geometries in terms of a probability distribution.
The simplest realization is given by the state that, while associating the same wave function to each GFT quantum, neglects the connectivity information, corresponding to an infinite superposition of states associated to disconnected tetrahedra. This corresponds to a coherent state of the GFT field operator.

An effective dynamics can be extracted by truncating the Schwinger-Dyson equation of the considered GFT model. The GFT field is then replaced by a collective condensate wavefunction which approximates the full quantum dynamics with the saddle point equation, that is a mean field approximation. This corresponds to the GFT equivalent of the Gross-Pitaevskii equation for a Bose condensate.\\

This program has been implemented in the specific case of a GFT formulation of the loop quantum gravity EPRL spin foam model \cite{Engle:2007wy} for Lorentzian four-dimensional  gravity. The simplicial combinatorics in the interactions is well defined, together with the use of GFT fields over $SU(2)^4$, and the associated embedding into a covariant $SL(2,\mathbb{C})$ symmetry. A free, massless, minimally coupled real scalar field is then added as the matter content. The resulting dynamics -- after several other natural approximations -- is given by the following generalized Friedmann equation:
\be
\f{V''}{V} = \f{2 \sum_j V_j \Big[ E_j + 2 m_j^2 \rho_j^2 \Big]}{\sum_j V_j \rho_j^2} \qquad ,
\ee
where $V$ is the volume of the Universe, $E_j$ are conserved quantities associated with the symmetries, $\rho_j$ are energy densities, and $m_j$ are related with the ratios of the constants entering the definition of the $E_j$. These equations have the correct classical limit. In addition, the requirement of a non-zero energy density for the scalar field must be assumed so as to have a correct interpretation in terms of a classical spacetime, implying that at least one of the $Q_j$ (conserved quantities) must be non-zero. This implies in turn that the volume remains positive with a single turning point, leading to the existence of a bounce replacing the big bang singularity. This is in agreement with other loop quantum gravity results. This is a key result for cosmology with possible observational signatures.\\

More realistic GFT condensate states, defined by superpositions of refined spin network graphs (triangulations), have also been successfully constructed. Some advances have also been made in the extension of condensate cosmology to the study of cosmological perturbations but the results are still preliminary. There are also conceptual issues around the notion of geometrogenesis and its physical interpretation as a fundamental dynamical process. The exhaustive and rigorous mathematical  characterization of realistic vacuum states of the full theory is also still an open question, together with the generic study of cosmological phase transitions in this framework. Although it is probably fair to emphasize that we are not close to having a full set of reliable cosmological predictions from GFT, it is already an impressive success that key cosmological features where recovered. The background dynamics may already lead to some specific observable consequences. This quantum gravity approach is therefore not far from establishing contact with experiments although clear non-ambiguous predictions are yet missing.

\section{Causal sets}

As argued by Dowker in \cite{Dowker:2017zqj}, the main bet of the causal set approach to quantum gravity is the fundamental discreteness of spacetime. This discreteness at the Planck scale is not expected to imply any major change in the well known low-energy technics used in standard physics. The causal order of spacetime is the basic concept in general relativity. Topology, differentiable structure and metric are probably less fundamental than the causal structure itself. This is strongly suggested by the fact that the causal order of a strongly causal  spacetime basically determines its chronological structure, its local null geodesics, its topology, its differentiable structure, and its conformal metric. In D=4, causal order is a unifying concept, encoding the spacetime geometry and missing only information about local physical scales. The other ingredient of the causal set approach is spacetime discreteness which is fundamental: the histories in the path integral for quantum gravity are discrete and no continuum limit is  taken in the full theory. Considering simultaneously causal order and discreteness straightforwardly leads to a \textit{discrete order}.\\

A causal set $(C, \prec)$ leads to an approximate continuum  because the order relation $\prec$ underpins the causal order
of $(M,g)$ and the physical scales that are missing are given by the ``atomicity", that is the number of elements 
in a portion of the causal set entering the spacetime volume of the region of the considered quasi-continuum. Geometry is given by numbers and order. In a causet -- the word causet is used as a shorthand for causal set -- $(C,\prec)$ with elements $x$ and $y$, if $x \prec y$ it is said that $x$ precedes $y$. If $x\ne y$ are unrelated by $\prec$ (which is written $x\natural y$), it is said that ``$x$ and $y$ are unrelated''. The order is irreflexive: $x\nprec x$. \\

An interesting successful prediction has been made for cosmology in this framework: the value of the cosmological constant has been correctly estimated before its measurement. Under the assumption that some dynamics drives $\Lambda$ towards zero and that the observed value is a fluctuation, it was suggested by Sorkin that 
$\Lambda \sim \Delta \Lambda \sim (\Delta V)^{-1}
\sim \pm \frac{1}{\sqrt{V}}\sim\pm 10^{-120} $ in Planck units (using the Hubble volume for $V$). The argument is not a rigorous proof from first principles but it remains a unique case of a (correct) cosmological prediction made by a quantum gravity theory.\\

Interesting concrete questions can also be asked in this context concerning Smolin's hypothesis of black holes leading to baby universes and to a cosmic natural selection paradigm. The question of whether this scenario can be realized  dynamically in the theory of causets is actually transformed into technical questions about the class of classical sequential growth  models. In particular one need to determine if there are partial breaks in causets grown in any classical sequential growth models and, if yes, what is the consequence of the renormalisation of the parameters entering the model. Although those questions remain open it is encouraging that the somehow vague initial statement can be translated into a precise proposal in the paradigm. \\

In addition, Sorkin's suggestion that a  large number of cycles of cosmic expansion and collapse might take place, punctuated by posts, could result in a dynamics for our current era such that the expansion after the latest post would have led to a  large, flat, D=3 space at the end of the Planck era. This is actively considered in the causet approach. 

\section{Asymptotic safety}

A known way out of the perturbative non-renormalizability of general relativity is to treat it as an effective field theory. This leads to a renormalizable gravitationnal theory if all possible counterterms fulfilling the symmetries are added in the action \cite{Weinberg:2009bg}. This works correctly at sub-Planckian energies because higher-derivative terms are suppressed by powers of the Planck mass, but using the approach at trans-Planckian energies would require to fix an infinite number of coupling constants. As explained in details in \cite{Bonanno:2017pkg}, Asymptotic Safety lives is an effective field theory which tries to resolve the ``predictivity'' problem usually encountered by imposing the following extra condition: the quantum theory describing our Universe should be located within the ultraviolet critical hypersurface of a suitable renormalization group (RG) fixed point. This addition means that the UV behavior of the theory is controlled by the fixed point which makes all dimensionless coupling constants finite at high energy.  

The approach makes sense because the existence of a RG fixed point has been shown and it was demonstrated that the UV critical hypersurface develops a regime where usual gravity is a correct approximation. As Asymptotic Safety is expected to be capable of describing gravity at large scales it seems suitable for cosmological model building. In addition, cosmological observations can be used to fix the free parameters entering the Asymptotic Safety construction. \\

A key ingredient in this approach is the functional renormalization group equation (FRGE) for the gravitational (effective averaged) action
$\Gamma_k$:
\be\label{FRGE}
\p_k \Gamma_k[g, \gb] = \frac{1}{2} {\rm Tr}\left[ \left( \Gamma_k^{(2)} + \cR_k \right)^{-1} \p_k \cR_k \right] \, . 
\ee
The flow equation agrees with the original hope of renormalizing by integrating out ``short scale perturbations'' with momenta $p^2 \ll k^2$. In this way, $\Gamma_k$ gives a valid description of physics for scales $k^2$. Importantly, the resulting RG flow dos not depend strongly on the background choice.

The simplest approximation of the RG flow consists in projecting the FRGE onto the Einstein-Hilbert action. Two scale-dependent coupling constants enter this game: the Newton constant $G_k$ and the cosmological constant $\Lambda_k$.
Detailed investigations show that this leads to two fixed points. Firstly, a Gaussian fixed point (GFP) which corresponds to a free theory whose stability coefficients are fixed through the mass-dimension of the coupling constant. Secondly, the flow possesses a non-Gaussian fixed point (NGFP) where the dimension of the Newton constant is anomalous. \\

Newton's constant and the cosmological constant are now scale-dependent at trans-Planckian energies. The question of the possible astrophysical consequences of this prediction then naturally arises. In the RG improvement process the cutoff can be identified with a caracteristic length scale in the system, $k(x^\mu)$. In the framework of cosmology, different types of cutoff can be considered:   
\begin{subequations}\label{cutoffids}
	\begin{align}
& \mbox{Type (I):} \quad	   &    k^2 & = \xi^2 \, t^{-2} \, ,  \label{typeI} \\
& \mbox{Type (II):} \quad   &	k^2 &= \xi^2 \, H(t)^2 \, , \label{typeII}\\
& \mbox{Type (III):} \quad  &	k^2 &= \xi^2 \,\sqrt{R_{\mu\nu\rho\sigma}R^{\mu\nu\rho\sigma}} \, ,  \label{typeIII}\\
& \mbox{Type (IV):} \quad   &	k^2 & = \xi^2 \,T^2 \, .  \label{typeIV}
	\end{align}
\end{subequations}

Supplementing the truncated Einstein equations by an equation of motion for the matter and using one of the cutoff suggestions given above leads to a closed system of equations which determines the averaged metrics $\langle g_{\mu\nu}\rangle_k$. The running of $\Lambda(k)$ and $G(k)$ leads to deviations of the dynamics with respect to the one of general relativity.
In a  way similar to what happens in QED or QCD, the RG approach to quantum gravity leads to improved actions by identifying $k$  with the strength of the field. This corresponds to  a cutoff  of Type (III) which leads to new terms in the equations of motion  originating from $D_\mu G(k(x)) \not = 0$.\\

The global strategy of RG improvement can be applied to the FLRW Universe and the resulting dynamics contains extra-terms corresponding to an energy transfer between the gravitational degrees of freedom and the matter sector. 

Interestingly it is possible to realize an inflationary phase in the fixed point regime with $\Omega_\Lambda^* \ge 1/2$ where inflation is driven by the quantum gravity effects and ends naturally when the RG flow tends to the classical regime. The NGFP-driven inflation may leave imprints in the CMB spectrum. The large-scale structures currently visible may have crossed the Hubble horizon during the NGFP regime. The power spectrum could then exhibit non-trivial corrections that are still hard to determine unambiguously.\\

Going deeper into the details requires the construction of effective actions. Most of them are of the $f(R)$ type. For example, a Type-(III) cutoff (and some additional hypotheses) leads to:
\be
\label{eff}
S=\frac{1}{2\kappa^2}\int d^4 x \sqrt{-g} \;\left [ R + \alpha R^{2-\frac{\theta_3}{2}}+ \frac{R^2}{6 m^2} -\Ls \right ] \, . 
\ee
The scalaron mass takes into account the details of the RG trajectory and  $\theta_3$ is the critical exponent of the $R^2$-operator (the $R^2$-coupling is encoded in $\alpha$).  The associated phenomenological parameters are not the same than the usual Starobinsky ones and it is possible that the next generation of experiments could distinguish between the models.

The asymptotic safety mechanism can be also operative in gravity-matter systems. The occurrence of non-Gaussian gravity-matter fixed points leads to dilaton-gravity (DG) models. A typical ansatz for the Euclidean effective average action is then
\be\label{dilatongrav}
\Gamma_k^{\rm DG} \approx \int d^4x \sqrt{g} \, \left[ \tfrac{1}{2} F_k(\chi^2) R - \tfrac{1}{2} K_k(\chi^2) \p_\mu \chi \, \p^\mu \chi - V_k(\chi) \right]  \, , 
\ee 
where the functions $F,V$ and $K$ depend on the scalar field $\chi$ and the RG scale $k$. Going to the Einstein frame allows to make predictions for the CMB that can then be calculated in the standard way.

\section{Causal dynamical triangulation}

Causal Dynamical Triangulation (CDT) provides a full non-perturbative quantum gravity proposal. The dynamics is obtained as the continuum limit of a regularized path integral. 
Some large-scale properties of the ground state of the universe can be explicitly calculated. Basically, CDT provides a regularized version of the standard
path integral over geometries $g$, the so-called ``sum over histories''
\begin{equation}
Z=\int {\cal D}g\, {\rm e}^{i S_{\rm EH}[g]}.
\label{formalPI}
\end{equation}

It was shown that in this approach, the continuum theory is mostly independent of the details of the regularization scheme. The  construction is rooted in the paradigm of quantum field theory, without requiring exotic ingredients. It is a kind of gravitational analogue of lattice QCD (although the symmetries are not the same). 
Beyond the linearized theory it is hard to implement consistently the full diffeomorphism symmetry, 
especially when the theory is discrete.
An important feature of CDT is that this problem is absent. This is because the configuration space of the regularized path integral is  
a space of triangulations, expressed by piecewise flat manifolds made of triangular building blocks of edge length $a$.
It is based on the  Regge approach to describe the curved spaces of GR by geometric data.

In standard CDT the  building blocks are arranged in layers labelled by a discrete time parameter $t$. The geometry of the spatial volume
for an integer value $t$, shared by two layers of ``height'' $\Delta t\! =\! 1$, 
is a flat Riemannian space made of equilateral tetrahedras with edge length $\ell_s$.

In addition, there exists in CDT a well-defined ``Wick rotation'', as an analytical continuation of the parameters describing the regularized flat geometries. It maps each 
curved Lorentzian CDT space-time to a curved Euclidean space-time and simultaneously transforms the complex
weight factor exp($iS$) to a real Boltzmann factor exp($-S_{\rm E}$).
There is no continuum analogue of this Wick rotation for general diffeomorphism equivalence classes of metrics.\\

The phase diagram of CDT as a quantum gravity proposal is spanned by three bare parameters. It is important to notice that CDT predicts a positive cosmological constant.
This feature has to do with the convergence of the non-perturbative path integral when the number of building blocks is taken to infinity.
However CDT does not predict any particular value for the physical $\Lambda$ of CDM cosmology.  \\

This phase diagram has been constructed thanks to Monte Carlo simulations.
Importantly, CDT has at least one second-order transition line -- this is a unique quantum gravity feature. The existence of such transition points 
is an important prerequisit to define the scaling limits of the theory. 
There is also a bifurcation phase characterized by the appearance of vertices of higher order, where many 4-simplices share the same vertex.
This might lead to a substructure of the (quantum) geometry, whose physical meaning is being tentatively 
associated with a possible breaking of spatial homogeneity.\\

Non-perturbative CDT shows that the usual ``sum over histories'' does not lead to a FLRW universe. The success of the FLRW description  
of the Universe at large scales is uneasy to fit in the quantum context where curvature fluctuations are large. They are associated with ``entropic'' instabilities, 
where the integral is dominated by degenerate configurations that don't average out to
lead to a correct semi-classical behavior. This is insensitive to the choice of the microscopic building
blocks.  The fundamental nature of these pathological ``spaces" is pre-geometric because they are dominated by
specific dynamical geometric modes, in a way such that a D=4 large universe is actually never formed. 

The main aim of CDT is precisely to tame the observed pathologies of
non-perturbative Euclidean dynamical triangulation models so as to allow a physical interpretation without
constraining  the local curvature. 

The main result is that requiring path integrals to have a well-behaved causal structure leads to a different quantum gravity theory. This causality condition is physically associated with the fact that the world is Lorentzian but is not a necessary requirement (as individual configurations are not intrinsically physical). 
This CDT hypothesis leads to one of the first explicit example of emergence of a specific cosmological space-time from a fundamental non-perturbative and 
background-independent quantum gravity model.\\

The fact that without assuming any preferred background geometry from the start, a
large region in the phase space of CDT leads to a  democratic superposition of all histories producing
a geometry with a shape matching the FLRW one is a strong result. This is also the case for the large-scale dimension
of the universe, which is predicted to be 4. There is a dynamical dimensional reduction to a value near 2
when one approaches the Planck scale. This has triggered a lot of interested and was also found in different approaches to quantum gravity. \\

However, the dynamically generated quantum space-time of CDT lives in a deeply non-perturbative regime and cannot be used as a classical de Sitter background.  
Usual approaches  cannot be imported to this non-perturbative setting, unless the associated observables
can be reformulated in a diffeomorphism-invariant and background-independent manner which is non-trivial.
Nevertheless, in CDT, these questions can be addressed in a well-defined framework.

The emergence of a dS space-time in CDT is genuinely non-perturbative and has no analogue in standard quantum cosmology. The (quantum) space-time built by the CDT 
path integral is the state which minimizes the Euclidean action determining the dynamics. This effective action receives important contributions from
both the bare action and the entropy. The associated phase transitions appear to be ``entropy-driven''. Interestingly, in the phase space region where a stable de Sitter vacuum state can be found, the CDT path integral does not exhibit any conformal-factor pathology (the divergence is present but entropically suppressed).  One can then obtain a dS non-perturbative ground state from the full quantum theory, without invoking a non-standard ``Wick rotation''.\\

At this stage the success of CDT quantum gravity relies more in being able to reproduce basic large scale cosmological features than on detailed specific predictions. This is however already a very nice and quite rare result from a full quantum gravity theory.

\section{Thermodynamics of spacetime}

From the pioneering work of Jacobson \cite{Jacobson:1995ab} to the recent study of Verlinde \cite{Verlinde:2016toy} many articles have been devoted to the idea of emergent gravity. I will here refer only to the specific ideas developed in \cite{Padmanabhan:2016lul} by Padmanabhan.\\

One should keep in mind that the expansion of the Universe actually depends on the  chosen coordinates. A geodesic observer will see the Friedmann metric (and the Schwarzschild metric) as expanding, while some non-geodesic observers can find both of them static. Let us consider for example the metric:
 
  \begin{eqnarray}
ds^2 &=& - c^2 dt^2 + \frac{4}{9} \left[ \frac{9GM}{2(x+ct)}\right]^{2/3} dx^2 \nonumber\\
&+& \left[ \frac{9GM}{2} ( x + ct)^2 \right]^{2/3} 
[d\theta^2+\sin^2\theta d\phi^2].
  \label{eqn7.52}
 \end{eqnarray} 
 
The area $A$ of the 2-surfaces with $t,x=$ constant increases with time as does the  volume enclosed by this surface. Observers using $t, x, \theta, \phi$ can claim that spacetime is expanding. But the metric of Eq. \ref{eqn7.52} actually describes the spacetime outside a spherical body like a star. Instead of the usual Schwarzschild coordinates, one just uses the coordinates describing observers in free fall. This is important to keep this is mind as some conundrums in the FLRW universe might seem to be related to the fact that spatial sections of the universe are expanding but this is not actually the case.\\

It should then be also reminded that the FLRW universe is under-determined in the sense that a relation between $p$ and $\rho$ must be chosen together with some conditions on acceptable behaviors for $p$ and $\rho$. Otherwise the model is not predictive as any function $a(t)$ (or $H(t)$) ca be obtained by a correct choice of the equation of state $p=p(\rho)$. In addition, if one starts with an equation of state known from laboratory experiments and evolve the universe backward in time, one reaches at some point an energy scale which is necessarily not tested. It means that it is impossible to work out fully a cosmological model relying on a lab-tested $T^a_b$ and then solve it for $g_{ab}$. Quite a lot of difficulties also arise from the fact that geometry does not constrain the matter sector enough in standard general relativity.\\

Some ``strange" numbers also enter the game. It is know that the cosmological constant density is an extraordinary small number:
  \begin{equation}
  (\rho_{\Lambda} L_P^4)\approx 1.1 \times 10^{-123}.
  \end{equation} 
It is probably less known that 
  \begin{equation}
  \rho_{\rm eq}L_P^4 \approx 2.5 \times 10^{-113},
  \end{equation}
where, so as to describe the radiation and matter dominated epochs, was introduced a \textit{constant} density 
  \begin{equation}
  \rho_{\rm eq}\equiv\frac{\rho_m^4(a)}{\rho_R^3(a)}=\sigma T_{eq}^4,
   \label{eqn4.9}
  \end{equation}
 the second equality defining the temperature $T_{\rm eq}$. The third important number is $\rho_{\rm inf}$ defined by $H^2(t)\approx (8\pi G\rho_{\rm inf})/3$ during inflation. 

The Universe is mostly described by three numbers $\rho_{\rm inf}, \rho_{\rm eq}, \rho_\Lambda$ that are not related either numerically or conceptually. Using only Einstein's equations is not enough to constrain the matter/radiation sector as any set of numbers ($\rho_{\rm inf}, \rho_{\rm eq}, \rho_\Lambda$) can in principle be obtained. Probably, an \textit{additional} theoretical principle is needed to make sense out of these numbers.
It is then convenient to introduce a dimensionless number $I$ liking those densities by:
\begin{equation}
I= \frac{1}{9\pi} \, \ln \left( \frac{4}{27} \frac{\rho_{\rm inf}^{3/2}}{\rho_\Lambda\,\rho_{\rm eq}^{1/2}}\right).
\label{strange1}
\end{equation} 
Numerically, one obtains
\begin{equation}
I   \approx 4\pi \left( 1 \pm \mathcal{O} \left(10^{-3}\right)\right).
\label{strange2}
\end{equation} 
It can actually be shown that this value can be understood as the amount of information accessible to an eternal observer and that this specific value is related with the quantum microstructure of spacetime.\\
 
It is also important to notice that at large scales the Universe selects a kind of preferred frame in which the CMB looks homogeneous and isotropic. This ``cosmic rest frame" is defined purely from observations, through the content of the Universe. The interesting paradox relies in the fact that the Friedmann model provides a fully covariant procedure to construct an absolute rest frame.

Moreover, one should also wonder why does the Universe expand (leading to an arrow of time) as Einstein's equations are time reversal invariant. How does this happen ? It is known that the scale factor $a(t)$ has the wrong sign in the kinetic energy term of the action and therefore corresponds an unstable mode. Some analogies with reversed harmonic oscillators do suggest that it is this  instability which we call expansion, and which, for times larger than the Planck time, picks up an arrow of time. This feature  occurs  in any effective theory framework describing semi-classical gravity once $a(t)$ acquires an unstable dynamics. This happens because all quantum cosmological models approach the usual Friedmann equations at times much larger than the Planck time.\\

For all those reasons, Padmanabhan advocates a change of paradigm. The dynamics of spacetime can be written in a purely thermodynamical language, relying on suitable degrees of freedom in the bulk and on the boundary. It was shown that the dynamics of geometry, understood as the heating and cooling of null surfaces, is given by:
\begin{equation}
\int_\mathcal{V} \frac{d^3x}{8\pi L_P^2}\sqrt{h} u_a g^{ij} \pounds_\xi p^a_{ij} = \epsilon\frac{1}{2} k_B T_{\rm avg} ( N_{\rm sur} - N_{\rm bulk}),
\label{evlnsnb}
\end{equation} 
with
\begin{equation}
 N_{\rm sur}\equiv\int_{\partial \mathcal{V}} \frac{\sqrt{\sigma}\, d^2 x}{L_P^2},\quad
N_{\rm bulk}\equiv \frac{|E|}{(1/2)k_BT_{\rm avg}},
\end{equation} 
 being the degrees of freedom on the surface $\partial \mathcal{V}$ and in the volume $\mathcal{V}$,  $T_{\rm avg}$ is the Unruh temperature of the boundary, and  $h_{ab}$ is the induced metric. This gives the gravitational dynamics  as a  thermal evolution of spacetime.
Importantly, although Eq. \ref{evlnsnb} is a time evolution, it is derived from an extremum condition for a thermodynamic variational principle and is associated with the thermodynamical equilibrium between degrees of freedom. A simpler version of this equation can easily be written in the cosmological setting. The question of whether the microscopic degrees of freedom have reached their maximum entropy state at the Hubble scale naturally arises.\\

The key ingredient needed to go ahead in this setting might be the concept of information stored in spacetime and its availability to different observers. This might lead to a relation between the current accelerated expansion and the early inflationary phase through the information content of spacetime. The concrete implementation of this idea leads to an interesting solution to the strange numbers mentioned above and suggests to interpret rigorously the $I$ quantity as the amount of information accessible to an eternal observer in our universe, as mentioned before. Its value is fixed by the effective 2-dimensionality of spacetime at the Planck scale.  The amount of information  accessible is reduced from an infinite amount to a finite value, called $I_c$, as a  consequence of the fact that $\Lambda \neq 0$. The detailed calculation of $I_c$ interestingly leads to a value of the cosmological constant which is in agreement with observations. This also predicts a correct amount of inflationary e-folds. Finally it is possible to calculate the energy scale of inflation ($1.2\times 10^{15}$ GeV) which make the model falsifiable. 
 
\section*{Acknowledgments}
I would like to thank Jacques Villain and the French ``Acad\'emie des sciences" for having invited me to coordinate this special issue on testing quantum gravity  with cosmology. 
\bibliography{refs}
\end{document}